\documentclass[pra,aps,amsmath,amssymb,showpacs]{revtex4}
\usepackage{graphicx, subfigure}
\usepackage{subfigure}
\usepackage{color}

\newcommand{\be}{\begin{equation}}
\newcommand{\ee}{\end{equation}}
\newcommand{\bea}{\begin{eqnarray}}
\newcommand{\eea}{\end{eqnarray}}
\newcommand{\ba}{\begin{array}}
\newcommand{\ea}{\begin{array}}

\begin{document}
\title{The  operator sum-difference  representation for  quantum maps:
  application to  the two-qubit amplitude  damping channel} \author{S.
  Omkar}  \email{omkar@poornaprajna.org} \affiliation{Poornaprajna Institute of  Scientific Research,
  Sadashivnagar,  Bengaluru-  560080,  India.}  \author{R.   Srikanth}
\email{srik@poornaprajna.org} \affiliation{Poornaprajna    Institute    of   Scientific    Research,
  Sadashivnagar,   Bengaluru-   560080,  India.}    \affiliation{Raman
  Research   Institute,  Sadashivnagar,   Bengaluru-   560060,  India}
\author{Subhashish   Banerjee}  \email{subhashish@iitj.ac.in} \affiliation{   Indian   Institute  of
  Technology Rajasthan, Jodhpur- 342011, India}

\begin{abstract}
On account of the Abel-Galois no-go theorem for the algebraic solution
to quintic  and higher order  polynomials, the eigenvalue  problem and
the associated characteristic equation for a general noise dynamics in
dimension $d$  via the Choi-Jamiolkowski approach cannot  be solved in
general  via  radicals.  We  provide  a  way  around this  impasse  by
decomposing the  Choi matrix  into simpler, not  necessarily positive,
Hermitian operators that are  diagonalizable via radicals, which yield
a set of `positive' and  `negative' Kraus operators.  The price to pay
is  that the  sufficient number  of Kraus  operators is  $d^4$ instead
of $d^2$, sufficient  in the  Kraus representation.  We  consider various
applications of the formalism:  the Kraus repesentation of the 2-qubit
amplitude damping  channel, the noise resulting from  a 2-qubit system
interacting dissipatively  with a vacuum bath;  defining the maximally
dephasing and  purely dephasing components  of the channel in  the new
representation, and studying their entanglement breaking and broadcast
properties.
\end{abstract}


\maketitle

\section{Introduction}

Any practical use of a  quantum operation involves taking into account
the effect  of the  ambient environment, and  the systematic  study of
such an  effect is called open  quantum systems.  This  now prevades a
vast arena of studies, see for e.g., the Refs. \cite{bp02,weiss} for two
distinct  flavors of  the  subject.  The  effect  of the  environment,
interchangeably called  the reservoir or the bath,  effects the system
dynamics, in general, in two  ways depending upon the commutability of
the system and  interaction Hamiltonian. If the two  commute, then the
process is  a quantum nondemolition  one, that is, there  is dephasing
without any energy exchange \cite{bg03}; while if they do not commute,
then  there is  dephasing along  with dissipation  \cite{bs07}.  These
effects have been brought within the ambit of practical implementation
by a number of very impressive experiments, for e.g., \cite{turchette}
involving  ion  traps and  \cite{brune}  using  high-Q cavity  quantum
electrodynamics.

As a  result, the use of  ideas from open quantum  systems have become
widespread in quantum information processing \cite{nc}.  A very useful
tool  in this regard  is the  Kraus representation  \cite{kraus} which
encodes  the effect  of the  environment  on the  system of  interest.
Since  a  quantum operation  that  can  be  represented in  the  Kraus
representation  is guaranteed  to  be completely  positive (CP),  it is  of
importance to  find the  Kraus representation pertaining  to different
open system  models. The main  aim of this  paper is to  develop Kraus
representations for general N-qubit open system models, and use it 
specifically on a two-qubit model much  discussed in
the  literature \cite{ft02}.   The  model considered  is  that of  two
qubits interacting with a bath,  for e.g., an electromagnetic field in
a squeezed thermal  state, via the dipole interaction,  which has been
considered in  detail for both  pure dephasing \cite{brs1} as  well as
dissipative    \cite{brs2}    system-reservoir   interactions.     The
system-reservoir coupling  constant is dependent upon  the position of
the  qubit, leading to  interesting dynamical  consequences. Basically
this allows  a classification  of the dynamics  into two  regimes: the
independent (or  localized) decoherence regime,  where the inter-qubit
distances  are such that  each qubit  sees an  individual bath  or the
collective decoherence  regime, where the  qubits are close  enough to
justify a collective interaction with the bath.

When  we  attempt  to  obtain   the  Kraus  operators,  for  both  the
independent  as  well as  collective  regimes  for  a version  of  the
dissipative model \cite{brs2}, described below, we encounter a problem
which has its origin in the famous Abel-Galois irreducibility theorem.
We are able  to circumvent this mathematical obstruction,  in the case
considered by resorting to the inherent symmetries in the model.  This
is  accomplished   by  introducing  the  concept   of  extended  Kraus
operators.  We can thus  coin the word \textit{Abel-Galois integrable}
for such  systems. It  would be  of interest, here,  to note  that the
label  \textit{Abel-Galois integrable} could  be ascribed  to infinite
dimensional systems also, such  as the dissipative harmonic oscillator
\cite{gsi}, which  due to their  inherent symmetries can be  solved by
quadratures.

The plan of  the paper is as follows. In the  next section, we briefly
discuss, the Abel-Galois irreducibility  theorem, due to its relevance
to  our work.   We then,  in anticipation  of its  need for  the Kraus
operators of  the general  two-qubit dissipative model,  introduce the
concept  of  extended  Kraus  operators.   This is  followed  by  some
illustrative applications  to a typical  single-qubit channel. The
use of this  formalism is, of course, not needed  for the single qubit
case, where it is possible to obtain the usual Kraus operators, but is
intended to serve  as an illustrative example to  the main application
of extended Kraus operators  to two-qubit dissipative interaction with
a vacuum bath,  the 2AD (amplitude damping) channel.  We then indicate
why this would  not work for the case of a  bath at finite temperature
and  bath  squeezing,  the  two-qubit squeezed  generalized  amplitude
damping channel  (2SGAD).  Next, we  discuss some features of  the 2AD
channel. Finally, we make our conclusions.

\section{Abel-Galois irreducibility theorem \label{sec:ag}}

The first famous no-go  result in algebra, the irreducibility theorem,
discovered independently by mathematicians  Niels Henrik Abel and Jean
Evariste  Gallois (and  anticipated  earlier by  Ruffini) states  that
polynomial equations of degree 5  or higher do not \textit{in general}
have solutions that can be  expressed algebraically, i.e., in terms of
addition, subtraction, multiplication, division  and taking roots to a
given  finite  order over  the  equation's  coefficients and  rational
numbers (and  any finite number of irrationals).   Obviously there are
infinitely many examples  where solutions thus expressible \textit{do}
exist, a  trivial example being $f_5(x)  \equiv (x - c)^5  = 0$, where
$c$  is a  real or  complex number,  and where  the solution  $x=c$ is
manifest.  Similarly solutions to a  product of a quartic and a linear
polynomial,  $f_4(x)f_1(x) =  0$,  can be  obtained  by solving  those
polynomials separately.  For the general  case of $n \ge 5$, one would
have to  resort to numerical methods  like the Laguerre  method or the
Newton-Raphson method.

Briefly,  the argument,  which is  part  of Galois  theory, runs  thus
\cite{art04}:  a polynomial  equation over  rational numbers  (or more
generally, over the  base field of given constants)  admits a solution
by   radicals   precisely   if   its  \textit{Galois   group}   is   a
\textit{solvable  group}.  For  polynomials upto  quartic  degree, the
associated  Galois  group is  solvable,  but  for  quintic and  above,
unsolvable cases exist.

Let the base field be the set $\mathbb{Q}$ of all rationals (augmented
by at  most a finite number  of irrational constants,  which we ignore
for simplicity). Suppose we are given a polynomial
\begin{equation}
f(x) = \sum_{j=0}^5 \alpha_jx^j \in \mathbb{Q}[x]. \label{eq:Q}
\end{equation}
If $x_k$ are the solutions to $f(x) = 0$, then:
\begin{equation}
f(x) = \Pi_{k=1}^5 (x - x_k) \in \mathbb{E}, \label{eq:E}
\end{equation}
where the  solutions $x_k$ exist  in general in  the \textit{splitting
  field} $\mathbb{E}$,  the extension field of  $\mathbb{Q}$, which is
the smallest subfield of $\mathbb{C}$ containing the roots of $f(x)$.

Multiplying   out  Eq.   (\ref{eq:E})  and   comparing  it   with  Eq.
(\ref{eq:Q})  shows  the   $\alpha_j$'s  to  be  elementary  symmetric
functions of the roots $x_j$:
\begin{eqnarray}
\beta_4 &=& \sum_j x_j;
\beta_3 = \sum_{j \ne k} x_j x_k;
\beta_2 = \sum_{j \ne k, j \ne l, k \ne l} x_j x_k x_l, \nonumber \\
\beta_1 &=& \sum_j \frac{x_1 x_2 x_3 x_4 x_5}{x_j};
\beta_0 = \Pi_j x_j,
\end{eqnarray}
where   $\alpha_j  =  (-1)^{j+1}\beta_j$.    We  observe   that  every
permutation $\sigma$ of the roots, which is clearly an automorphism of
$\mathbb{E}$, leaves $\beta_j \in \mathbb{Q}$, and hence $\mathbb{Q}$,
fixed.   It  can  be shown  that  this  is  the only  automorphism  of
$\mathbb{E}$   that   leaves  $\mathbb{Q}$   fixed.    The  group   of
automorphisms $\sigma$  of $\mathbb{E}$ such  that $\sigma(q)=q$, with
$q  \in \mathbb{Q}$  is called  the Galois  group associated  with the
polynomial,  and denoted Gal$(\mathbb{E}/\mathbb{Q})  \subseteq S_n$,
the group of permutations on the roots of $f(x)$.

A subgroup $N$ of group $G$ is  called normal if and only if $gN = Ng$
for all $g \in G$, or equivalently, $g^{-1}Ng = N$, that is the normal
subgroup  is invariant  under  conjugation of  elements  of $G$.   The
relationship  is  denoted $N  \triangleleft  G$.  The normal  subgroup
defines  the quotient  or factor  group, denoted  $G/N$.   A subnormal
series (or tower) of a group $G$ is a sequence of subgroups:
\begin{equation}
\{e\}  \triangleleft  A_0 \triangleleft  A_1  \triangleleft A_2  \cdots
\triangleleft A_n = G,
\end{equation}
terminating in the trivial  subgroup. A \textit{composition series} is
a subnormal tower such that $A_j \ne A_{j+1}$ and $A_j$ is the maximal
normal subgroup  of $A_{j+1}$. The group  $G$ is solvable  only if the
factor groups $A_{j+1}/A_j$ in the composition series are Abelian. The
Abel-Galois theorem can be stated as follows: a polynomial equation is
solvable by radicals iff its Galois group is solvable.

For  all $n$,  the  maximal normal  subgroup  of $S_n$  is $A_n$,  the
alternating group, which is the  subgroup of even permutations on $S$.
For $n  \le 4$, all subgroups  of $S_n$ are solvable.   However for $n
\ge 5$, $A_n$  is non-Abelian, as is the  factor group $A_n/\{e\}$, so
that  $S_n$ is not  solvable, and  we find  that polynomials  that are
quintic or of  higher degrees are in general  not solvable.

The significance of the Abel-Galois theorem for us is that our work on
deriving the Kraus operators, for general two-qubit systems, 
requires diagonalizing a (density) matrix
in a Hilbert  space of dimension $d^2=16$, which  involves solving the
characteristic equation  of a self-adjoint matrix of  degree $16$.  In
general,   the   relevant  Galois   group   is   $S_{16}$,  which   is
unsolvable. We  circumvent this problem,  in any dimension,  by making
use of symmetry properties of  the matrix to circumvent the problem of
diagonalization. At times, it may happen that even if the Galois group
is solvable, diagonalization can be so tedious that our alternative is
preferable.

\section{Extended Kraus operators \label{sec:exten}}

A transformation  of a quantum  state is  a CP map  if and only  if it
evolves a density operator according to the prescription \cite{choi}:
\begin{equation}
\rho \longrightarrow \rho^\prime = 
\sum_j A_j \rho A^\dag_j,
\label{eq:kraus}
\end{equation}
where $A_j$ are at most  $d^2$ operators that satisfy the completeness
condition $\sum_j A_j^\dag  A_j = \mathbb{I}$. Here we  will study the
conditions under which a CP map allows a description of the form:
\begin{equation}
\rho \longrightarrow \mathcal{E}(\rho) = \rho^\prime = 
\sum_{j=1}^\mu A^+_j \rho \left(A^+\right)^\dag_j -
\sum_{k=1}^\nu A^-_k \rho \left(A^-\right)^\dag_k,
\label{eq:kraus0}
\end{equation}
where $\mu +  \nu \ge d^2$ and the  extended Kraus operators $A_j^\pm$
must satisfy the new completeness condition
\begin{equation}
\sum_{j=1}^\mu  \left(A^+\right)^\dag_j A^+_j -
\sum_{k=1}^\nu  \left(A^-\right)^\dag_k A^-_k = \mathbb{I}.
\label{eq:complete}
\end{equation}
Let  $|\mathcal{A}_j^\pm\rangle$  represent  a vector  `unfolding'  of
$A^\pm_j$.  Then  the  necessary  and  sufficient  condition  for  the
extended operator sum representation to be CP is that:
\begin{equation}
\sum_{j=1}^\mu  |\mathcal{A}^+_j\rangle \langle\mathcal{A}_j^+|  -
\sum_{k=1}^\nu  |\mathcal{A}^-_k\rangle\langle\mathcal{A}_k^-|  
\equiv \mathcal{B^+} - \mathcal{B^-} =
\mathcal{B},
\label{eq:sum+-}
\end{equation}
where  $\mathcal{B}$   is  the  Choi  matrix  for   the  map. \color{black}

To  see this,  we note  that the  Choi matrix  corresponding  to noise
$\mathcal{E}$ can be written as:
\begin{equation}
\mathcal{B}   =   \sum_{jk}     |j\rangle\langle   k|   \otimes
\mathcal{E}\left(|j\rangle\langle k|\right).
\label{eq:choi}
\end{equation}
Let $\mathcal{B} = \sum_j \mathcal{B}_j$ be any Hermitian partition of
$\mathcal{B}$.   The  partition   elements   $\mathcal{B}_j$  can   be
spectrally  decomposed  with positive  or  negative eigenvalues.   Let
their corresponding eigenvectors, normalized  to the absolute value of
the      eigenvalues       be,      $|\mathcal{A}_i^+\rangle$      and
$|\mathcal{A}^-_i\rangle$,   respectively.    It   follows  from   the
properties of matrices that \cite{debbie}:
\begin{equation}
|\mathcal{A}^\pm_i\rangle\langle   \mathcal{A}^\pm_i|   =   \sum_{j,k}
\left(|j\rangle\langle   k|   \otimes  \left(A^{\pm}_i|j\rangle\langle
k|A^{\pm\dag}_i\right)\right).
\label{eq:choix}
\end{equation}
Inserting  these  values in  Eq.  (\ref{eq:sum+-})  and comparing  the
result with the expression in Eq. (\ref{eq:choi}), we have:
\begin{eqnarray}
\mathcal{B} &=& 
\sum_{j=1}^\mu |\mathcal{A}^+_j\rangle \langle\mathcal{A}_j^+|    -
\sum_{k=1}^\nu   |\mathcal{A}^-_k\rangle\langle\mathcal{A}_k^-| \nonumber \\
& = & 
\sum_{j,k} \left(|j\rangle\langle k| \otimes 
\left[\sum_{i=1}^{\mu} A^{+}_i\left(|j\rangle\langle k|A^{+\dag}_i\right)\right] - 
\left[\sum_{i=1}^{\nu} A^{-}_i\left(|j\rangle\langle k|A^{-\dag}_i\right)\right]\right) \nonumber \\
& = &\sum_{jk} |j\rangle\langle k| \otimes \mathcal{E}\left(|j\rangle\langle k|\right),
\label{eq:combo}
\end{eqnarray}
from which  Eq. (\ref{eq:kraus0}) follows,  because the sum-difference
operation on each element  $|j\rangle\langle k|$ reproduces the effect
of  $\mathcal{E}$  on  that  element,  and thus  also  on  any  linear
combination, such as $\rho$, of those elements.


We consider an  example of the above idea,  applied to the generalized
amplitude  damping   channel  (GAD)  \cite{bs07}.    The  Choi  matrix
corresponding          to           GAD          channel          with
elements $\left[\sqrt{p}\begin{pmatrix}1&0\\0&\sqrt{1-\lambda}\end{pmatrix},
  \sqrt{p}\begin{pmatrix}0&0\\           \sqrt{\lambda}&0\end{pmatrix},
    \sqrt{1-p}\begin{pmatrix}\sqrt{1-\lambda}&0\\0&1\end{pmatrix},
      \sqrt{1-p}\begin{pmatrix}0&\sqrt{\lambda}\\0&0\end{pmatrix}\right]$
is \be \mathcal{B}=
\begin{pmatrix}
1-\lambda+p\lambda&0&0&\sqrt{1-\lambda}\\
0&p\lambda&0&0\\
0&0&(1-p)\lambda&0\\
\sqrt{1-\lambda}&0&0&1-p\lambda
\end{pmatrix}.
\ee

A decomposition of the above matrix is
$\mathcal{B} = \mathcal{B}^+ - \mathcal{B}^-$, with
\be
\mathcal{B}^{+}=
\begin{pmatrix}
1-\lambda+p\lambda+\frac{\sqrt{1-\lambda}}{2}&0&0&\frac{3\sqrt{1-\lambda}}{4}\\ 0&p\lambda&0&0\\ 0&0&(1-p)\lambda&0\\ \frac{3\sqrt{1-\lambda}}{4}&0&0&1-p\lambda+\frac{\sqrt{1-\lambda}}{2}
\end{pmatrix};
\mathcal{B}^{-}=
\begin{pmatrix}
\frac{\sqrt{1-\lambda}}{2}&0&0&-\frac{\sqrt{1-\lambda}}{4}\\
0&0&0&0\\
0&0&0&0\\
-\frac{\sqrt{1-\lambda}}{4}&0&0&\frac{\sqrt{1-\lambda}}{2}
\end{pmatrix}.
\ee
Both the  matrices are seen to be Hermitian. The positive Kraus
operators                           are                           \bea
K_1^{+}&=&\frac{\sqrt{4+2\sqrt{1-\lambda}-2\lambda-\sqrt{a}}}{2}
\begin{pmatrix}-\frac{2\lambda(1-2p)+\sqrt{a}}{3\sqrt{1-\lambda}}&0\\0&1\end{pmatrix},\\ \nonumber 
K_2^{+}&=&\frac{\sqrt{4+2\sqrt{1-\lambda}-2\lambda+\sqrt{a}}}{2}
\begin{pmatrix}-\frac{2\lambda(1-2p)-\sqrt{a}}{3\sqrt{1-\lambda}}&0\\0&1\end{pmatrix},\\ \nonumber 
K_3^{+}&=&\begin{pmatrix}0&\sqrt{(1-p)\lambda}\\0&0\end{pmatrix}, \hspace{1cm}
  K_4^{+}=\begin{pmatrix}0&0\\\sqrt{p\lambda}&0\end{pmatrix},      \eea
  where   $a=9(1-\lambda)+4\lambda^2(1-2p)^2$.   The  negative   Kraus
  operators      are      \be      K_1^{-}=\frac{(1-\lambda)^{1/4}}{2}
  I, \hspace{1cm} K_2^{-}=\frac{\sqrt{3}(1-\lambda)^{1/4}}{2}\sigma_z.
  \ee

Of course,  the extended Kraus  formalism used here is  not necessary,
since the  usual Kraus  operators for these  channels can  be obtained
analytically. The purpose here was to serve 
as a simple example that sets the
scene,  in  what  follows,  for  the non-trivial  application  of  the
formalism in the two-qubit case.
 
\section{Introduction to two-qubit dynamics}

Consider  the Hamiltonian, describing  the dissipative  interaction of
two qubits with the bath via the dipole interaction as \cite{ft02}
\begin{eqnarray}
H & = & H_S + H_R + H_{SR} \nonumber \\ & = & \sum\limits_{n=1}^2 \hbar 
\omega_n S^z_n + \sum\limits_{\vec{k}s} \hbar \omega_k (b^{\dagger}_{\vec{k}s}
b_{\vec{k}s} + {1 / 2}) - i\hbar \sum\limits_{\vec{k}s}
\sum\limits_{n=1}^2 [\vec{\mu}_n . \vec{g}_{\vec{k}s} (\vec{r}_n)(S_n^+ + 
S_n^-)b_{\vec{k}s}- h.c.]. \label{hamiltonian} 
\end{eqnarray}
Here $H_S$ is the system, $H_R$ the bath, and $H_{SR}$ the interaction
Hamiltonians,  respectively,  and  $\vec{\mu}_n$  are  the  transition
dipole   moments,  dependent   on  the   different   atomic  positions
$\vec{r}_n$.  Also,  $S_n^{\pm}$ are  the dipole raising  and lowering
operators, respectively  while $S_n^z$ is  the energy operator  of the
$n$th  atom, and  $b^{\dagger}_{\vec{k}s}$,  $b_{\vec{k}s}$ are  the
creation  and   annihilation  operators  of  the   (bath)  field  mode
$\vec{k}s$ with  the wave  vector $\vec{k}$, frequency  $\omega_k$ and
polarization index $s =  1,2$ with the system-reservoir (S-R) coupling
constant
\begin{equation}
\vec{g}_{\vec{k}s} (\vec{r}_n) = (\frac{\omega_k}{2 \varepsilon \hbar V})^
{1/2} \vec{e}_{\vec{k}s} e^{i \vec{k}.r_n}. \label{coupling}
\end{equation}
In  Eq.   (\ref{coupling})  $V$   is  the  normalization   volume  and
$\vec{e}_{\vec{k}s}$ is the unit polarization vector of the field.  It
can be seen  from Eq. (\ref{coupling}) that the  S-R coupling constant
is dependent on  the atomic position $r_n$ leading  to the possibility
of considering the dynamics  in the independent or collective regimes,
depending on whether the qubits are far apart or close with respect to
the length scales set  by the environment.  Assuming separable initial
conditions,  and taking  a trace  over  the bath  the reduced  density
matrix of the  two-qubit system can be obtained  \cite{brs2}.  We will
now attempt to obtain the Kraus operators for this model, in a unified
way for both the independent  as well as collective regimes, first for
the case  of a  vacuum bath, that  is, the  2AD channel and  then, 
point out the difficulties when we encounter finite temperature and bath squeezing.

The two-qubit density matrix in dressed state basis is
\be
 \rho= \left(
\begin{array}{clclr}
\rho_{ee}&\rho_{es}&\rho_{ea}&\rho_{eg}\\
\rho_{se}&\rho_{ss}&\rho_{sa}&\rho_{sg}\\
\rho_{ae} & \rho_{as}&\rho_{aa}&\rho_{ag}\\
\rho_{ge} & \rho_{gs}&\rho_{ga}&\rho_{gg}
\end{array}
\right).
\ee 
The  time-evolution   of  the  two-qubit  density   matrix  $\rho$  to
$\rho^\prime = \mathcal{E}(\rho)$ is given by:
\be 
\rho^\prime =
\left(
\begin{array}{clclr}
A\rho_{ee}&J\rho_{es}&M\rho_{ea}&L\rho_{eg}\\
J^\ast\rho_{se}&B\rho_{ss}+C\rho_{ee}&P\rho_{sa}&T\rho_{sg}+(U+iV)\rho_{es}\\
M^\ast\rho_{ae} & P^\ast\rho_{as}&D\rho_{aa}+E\rho_{ee}&Q\rho_{ag}+(iS-R)\rho_{ea}\\
L^\ast\rho_{ge} & T^\ast\rho_{gs}+(U^\ast-iV^\ast)\rho_{se}&Q^\ast\rho_{ga}+(-iS^\ast-R^\ast)\rho_{ae}&
\rho_{gg}+F\rho_{ss}+G\rho_{aa}+H\rho_{ee}
\end{array}
\right),
\label{eq:channel}
\ee
where $A,B,C,\dots, U,V,S,R$ are given Appendix \ref{sec:a}. 

The  Choi matrix  for  the  above interaction,  which  is the  density
operator  $(\mathcal{E}\otimes \mathbb{I}) (|\Phi\rangle\langle\Phi|)$,
where $|\Phi\rangle \equiv |00\rangle|00\rangle + |01\rangle|01\rangle
+  |10\rangle|10\rangle  +  |11\rangle|11\rangle$,  is given  by:  \be
\mathcal{B} =  \left(\begin{array}{cccccccccccccccc} A&0&0&0 & 0&J&0&0
  &   0&0&M&0   &   0&0&0&L\\   0&C&0&0   &  0&0&0&U+iV   &   0&0&0&0   &
  0&0&0&0\\ 0&0&E&0 & 0&0&0&0 &  0&0&0&iS-R & 0&0&0&0\\ 0&0&0&H & 0&0&0&0
  & 0&0&0&0 & 0&0&0&0\\

0&0&0&0 & 0&0&0&0 & 0&0&0&0 & 0&0&0&0\\
J^\ast&0&0&0 & 0&B&0&0 & 0&0&P&0 & 0&0&0&T\\
0&0&0&0 & 0&0&0&0 & 0&0&0&0 & 0&0&0&0\\
0&0&U^\ast-iV^\ast&0 & 0&0&0&F & 0&0&0&0 & 0&0&0&0\\

0&0&0&0 & 0&0&0&0 & 0&0&0&0 & 0&0&0&0\\
0&0&0&0 & 0&0&0&0 & 0&0&0&0 & 0&0&0&0\\
M^\ast&0&0&0 & 0&P^\ast&0&0 & 0&0&D&0 & 0&0&0&Q\\
0&0&-iS^\ast-R^\ast&0 & 0&0&0&0 & 0&0&0&G & 0&0&0&0\\

0&0&0&0 & 0&0&0&0 & 0&0&0&0 & 0&0&0&0\\
0&0&0&0 & 0&0&0&0 & 0&0&0&0 & 0&0&0&0\\
0&0&0&0 & 0&0&0&0 & 0&0&0&0 & 0&0&0&0\\
L^\ast&0&0&0 & 0&T^\ast&0&0 & 0&0&Q^\ast&0 & 0&0&0&1
\end{array}\right).
\label{eq:pma}
\ee It turns out (as can be found using Mathematica software) that the
relevant characteristic equation is cubic, so that the Galois group is
a  subset   of  $S_3$.   This  is   so  because  the   matrix  in  Eq.
(\ref{eq:pma}) is sparse,  and is found not to  be the case otherwise.
However the solutions are so  tediously long that they would be hardly
of practical interest, and it is advantageous to use our method, which
is   applicable  quite  generally   (even  for   non-sparse  Hermitian
matrices).

As a result, the formalism of Section (\ref{sec:exten}) will be 
used to derive
extended Kraus operators. We find the following
decomposition convenient:
\begin{equation} 
\mathcal{B} \equiv \mathcal{B}_{\rm diag}
+ \mathcal{B}_J +  \mathcal{B}_M + \mathcal{B}_L + \mathcal{B}_P 
+ \mathcal{B}_Q +
\mathcal{B}_T   +   \mathcal{B}_U  +\mathcal{B}_V  +   \mathcal{B}_R +\mathcal{B}_S,
\label{eq:mathcalB}
\end{equation}
where $\mathcal{B}_{\rm diag}$ is the submatrix of
$\mathcal{B}$ consisting of precisely
the diagonal entries in $\mathcal{B}$, and 0's elsewhere;
$\mathcal{B}_J$ is the submatrix consisting only of the
conjugate terms $J, J^\ast$, and 0's elsewhere, and so on.  

\subsection{Diagonal terms}
It is straightforward to see that by diagonalizing $ \mathcal{B}_{diag}=(A|0000\rangle\langle 0000|+C|0001\rangle\langle 0001|+E|0010\rangle\langle 0010|+H|0011\rangle\langle 0011|+B|0101\rangle\langle 0101|+F|0111\rangle\langle 0111|+D|1010\rangle\langle 1010|+G|1011\rangle\langle 1011|+|1111\rangle\langle 1111|)$, we have
\be  \mathcal{B}_{diag}  \equiv
\sum_{j=1}^9 |{\mathcal K}^+_j\rangle\langle{\mathcal K}^+_j|.
\ee
Then the Kraus operators obtained by `folding' each eigenvector
$\mathcal{K}^+_j$ is:
\bea
K_H^{+}&=&\sqrt{H}\begin{pmatrix}
0&0&0&0\\
0&0&0&0\\
0&0&0&0\\
1&0&0&0
\end{pmatrix},~
K_G^{+}=\sqrt{G}\begin{pmatrix}
0&0&0&0\\
0&0&0&0\\
0&0&0&0\\
0&0&1&0
\end{pmatrix},~
K_F^{+}=\sqrt{F}\begin{pmatrix}
0&0&0&0\\
0&0&0&0\\
0&0&0&0\\
0&1&0&0
\end{pmatrix},\\ \nonumber
K_E^{+}&=&\sqrt{E}\begin{pmatrix}
0&0&0&0\\
0&0&0&0\\
1&0&0&0\\
0&0&0&0
\end{pmatrix},~
K_D^{+}=\sqrt{D}\begin{pmatrix}
0&0&0&0\\
0&0&0&0\\
0&0&1&0\\
0&0&0&0
\end{pmatrix},~
K_C^{+}=\sqrt{C}\begin{pmatrix}
0&0&0&0\\
1&0&0&0\\
0&0&0&0\\
0&0&0&0
\end{pmatrix},\\ \nonumber
K_A^{+}&=&\sqrt{A}\begin{pmatrix}
1&0&0&0\\
0&0&0&0\\
0&0&0&0\\
0&0&0&0
\end{pmatrix}, ~
K_1^{+}=\begin{pmatrix}
0&0&0&0\\
0&0&0&0\\
0&0&0&0\\
0&0&0&1
\end{pmatrix}, ~
K_B^{+}=\sqrt{B}\begin{pmatrix}
0&0&0&0\\
0&1&0&0\\
0&0&0&0\\
0&0&0&0
\end{pmatrix}. \label{diagonalkraus}
\eea
These operators lead to the evolution $\mathcal{E}_{\rm diag}$, 
which transforms only the diagonal elements, killing off the rest:
\be
\rho^\prime_{\rm diag} \equiv \mathcal{E}_{\rm diag}(\rho) =
\sum_{j \in \textbf{T}} K^+_j \rho \left(K^+_j\right)^\dag =
\left(
\begin{array}{clclr}
A\rho_{ee}&0&0&0\\
0&B\rho_{ss}+C\rho_{ee}&0&0\\
0 & 0&D\rho_{aa}+E\rho_{ee}&0\\
0 &0&0&\rho_{gg}+F\rho_{ss}+G\rho_{aa}+H\rho_{ee}
\end{array}
\right),
\label{eq:diag}
\ee
$j \in \textbf{T} \equiv \{H, G, F, E, D, C, A, B, 1\}$.

\subsection{Off-diagonal terms}

By diagonalizing $\mathcal{B}_J=J|0000\rangle\langle 0100|+J^{\ast}|0100\rangle\langle 0000|$, we have: \be
\label{eq:rhoj}
\mathcal{B}_J \equiv
|{\mathcal K}^+_{J}\rangle\langle{\mathcal K}^+_{J}| -
|{\mathcal K}^-_{J}\rangle\langle{\mathcal K}^-_{J}|.
\ee
By `folding' the vectors, we obtain the Kraus operators
\be
K_{J}^{-}=\sqrt{\frac{|J|}{2}}
\begin{pmatrix}
1&0&0&0\\
0&-e^{i\phi_J}&0&0\\
0&0&0&0\\
0&0&0&0
\end{pmatrix},~
K_{J}^{+}=\sqrt{\frac{|J|}{2}}
\begin{pmatrix}
1&0&0&0\\
0&e^{i\phi_J}&0&0\\
0&0&0&0\\
0&0&0&0
\end{pmatrix},
\ee
where $\phi_J=  -(\omega_0-\Omega_{12})t$.
 The evolution produced  by these operators transforms 
two conjugate  elements of $\rho$ into the  two corresponding elements
of $\rho^\prime$, while annihilating all other elements in the density
operator.    Thus: 
 \be  \rho^\prime_J   =  K_{J}^{+}\rho
(K_{J}^{+})^{\dagger}- K_{J}^{-}\rho \left(K_{10}^{-}\right)^{\dagger}
= \begin{pmatrix} 0 &e^{i\phi_J}|J|\rho_{es}&0&0\\ e^{-i\phi_J}|J|\rho_{se}& 0
  &0&0\\ 0&0&0&0\\ 0&0&0&0
\end{pmatrix}.
\label{eq:rhoJ}
\ee

Proceeding thus with the other terms in Eq. (\ref{eq:mathcalB}), we obtain
pairs of positive and negative Kraus operators:
\bea
K_{M}^{\pm}&=&\sqrt{\frac{|M|}{2}}
\begin{pmatrix}
1&0&0&0\\
0&0&0&0\\
0&0&\pm e^{i\phi_M}&0\\
0&0&0&0
\end{pmatrix},~
K_{L}^{\pm}=\sqrt{\frac{|L|}{2}}
\begin{pmatrix}
1&0&0&0\\
0&0&0&0\\
0&0&0&0\\
0&0&0& \pm e^{i\phi_L}
\end{pmatrix}, 
K_{P}^{\pm}=\sqrt{\frac{|P|}{2}}
\begin{pmatrix}
0&0&0&0\\
0&1&0&0\\
0&0&\pm e^{i\phi_P}&0\\
0&0&0&0
\end{pmatrix},\nonumber \\
K_{T}^{\pm}&=&\sqrt{\frac{|T|}{2}}
\begin{pmatrix}
0&0&0&0\\
0&1&0&0\\
0&0&0&0\\
0&0&0&\pm e^{i\phi_T}
\end{pmatrix},~
K_{U}^{\pm}=\sqrt{\frac{|U|}{2}}
\begin{pmatrix}
0&0&0&0\\
1&0&0&0\\
0&0&0&0\\
0&\pm e^{i\phi_U}&0&0
\end{pmatrix},~
K_{V}^{\pm}=\sqrt{\frac{|V|}{2}}
\begin{pmatrix}
0&0&0&0\\
1&0&0&0\\
0&0&0&0\\
0&\pm e^{i\phi_V}&0&0
\end{pmatrix},\nonumber \\
K_{Q}^{\pm}&=&\sqrt{\frac{|Q|}{2}}
\begin{pmatrix}
0&0&0&0\\
0&0&0&0\\
0&0&1&0\\
0&0&0&\pm e^{i\phi_Q}
\end{pmatrix},~
K_{S}^{\pm}=\sqrt{\frac{|S|}{2}}
\begin{pmatrix}
0&0&0&0\\
0&0&0&0\\
1&0&0&0\\
0&0&\pm e^{i\phi_S}&0
\end{pmatrix},~
K_{R}^{\pm}=\sqrt{\frac{|R|}{2}}
\begin{pmatrix}
0&0&0&0\\
0&0&0&0\\
1&0&0&0\\
0&0&\pm e^{i\phi_R}&0
\end{pmatrix},
\eea
where $\phi_L= -2\omega_0 t, \phi_M= -(\omega_0+\Omega_{12})t, \phi_P= -2\Omega_{12}t, \phi_T= -(\omega_0+\Omega_{12})t = \phi_U,
\phi_V= -(\omega_0+\Omega_{12})t + \pi/2,\phi_Q= -(\omega_0-\Omega_{12})t, \phi_S= -(\omega_0-\Omega_{12})t + \pi/2, \phi_R= -(\omega_0-\Omega_{12})t + \pi$.\\
These operators lead, analogously to Eq. (\ref{eq:rhoJ}) to the
partial evolutions $\rho^\prime_M, \rho^\prime_L, 
\rho^\prime_P, \rho^\prime_T, \rho^\prime_X, \rho^\prime_Q$ 
and $\rho^\prime_Y$, which satisfy:
\be
\rho^\prime = \rho^\prime_{\rm diag} +
\rho^\prime_J+\rho^\prime_M+\rho^\prime_L
+\rho^\prime_P+\rho^\prime_T+\rho^\prime_U+\rho^\prime_V+\rho^\prime_Q+
\rho^\prime_R+\rho^\prime_S
,
\ee
while
\begin{eqnarray}
\mathcal{B}   &=&   \sum_{j=1}^9
|\mathcal{K}_j^+\rangle\langle\mathcal{K}_j^+|    +   
\sum_{j \in \textbf{S}}
\left(         |\mathcal{K}_j^+\rangle\langle\mathcal{K}_j^+|        -
|\mathcal{K}_j^-\rangle\langle\mathcal{K}_j^-|\right),
\label{eq:kr}
\end{eqnarray}
where $\mathbf{S} = \{J, M, L, P, T, U,V, Q, R,S\}$.

In short, our strategy to circumvent the Abel-Galois theorem is
to replace the problem of diagonalizing the Choi matrix, by 
that of diagonalizing simpler Hermitian matrices that sum up
to the Choi matrix. The elements of this Hermitian partition
are diagonal or rank-two matrices, and thus trivially or readily
diagonalized, in the case we considered.
The price to pay is that the number of Kraus operators required
to implement the CP map on a density operator
in an $d$-dimensional Hilbert space can be as many as
$d^4 = d^2 + \frac{d^2(d^2-1)}{2} \times 2$
(cf. Section \ref{sec:agn}), which is in general 
more than $d^2$, the sufficient number of 
Kraus operators in the conventional formalism \cite{nc}.

\subsection{Abel-Galois non-integrability \label{sec:agn}}

In our method, $\mathcal{B}_{\rm diag}$
is a diagonal matrix
while $\mathcal{B}_i$ $(i=J, M, L, P,
Q,  T,  U,V,R,S)$ are  all  rank-2  Hermitian  matrices, which  are  readily
diagonalizable.  
We will assume  that the base field is $\mathbb{Q}(\mathcal{B})$,
i.e., $\mathbb{Q}$ augmented by the entries in $\mathcal{B}$. In that case,
the Galois group of $\mathcal{B}_{\rm diag}$ is trivial
(consisting  of  the   identity  permutation) in
that the splitting field of the characteristic equation 
\begin{equation}
\chi_{\rm diag} \equiv (x - A)(x - C)(x-E)(x-H)(x-B)(x-F)(x-D)(x-G)(x-1),
\end{equation}
is the base field itself.  
The ten remaining  component matrices of $\mathcal{B}$ 
are rank-2  matrices.   The  characteristic equation,  for
example, for $\mathcal{B}_J$ in Eq. (\ref{eq:rhoj}), is:
\begin{equation}
\chi_J \equiv x^2 - |J|^2,
\end{equation}
with  solutions $\pm |J| = \pm \sqrt{J_R^2 + J_I^2}$, which are
in general irrational. Hence the symmetry group is $S_2$
consisting of the identity element and an interchange.

Thus our method of circumventing the Abel-Galois theorem can be
considered as reducing a problem requiring the solution of
(the unsolvable) $S_{d^2}$ to one requiring that of 
$\left(S_2\right)^{\times d^2(d^2-1)/2}$.
It turns out that for the particular form (\ref{eq:pma})
of the Choi matrix for the 2AD channel, the characteristic
equation is cubic and indeed solvable (as can be found using
Mathematica).  However, the resulting eigenvalues and eigenvectors
are so ponderous, that our method is preferable.
 
For the two-qubit squeezed generalized amplitude damping (2SGAD) channel \cite{brs2},
the  problem is seen to be \textit{Abel-Galois non-integrable},
i.e., it does not admit analytic Kraus operators.

\section{An application of the operator sum-difference formalism}

We indicate some features of the operator sum-difference formalism
in identifying special cases of a given noise that may have special
interest.

\subsection{The maximally dephasing component}

It may be  interesting to note that the  Kraus operators $K_j^{+}$, $j
\in \textbf{T} \equiv \{H, G, F, E, D, C, A, B, 1\}$, which correspond
to the diagonal  terms in the Choi matrix,  by themselves constitute a
CP   trace-preserving  (CPT)   dynamics.   Physically,   this  channel
represents  a  dynamics of  the  populations  (diagonal  terms of  the
density operator)  that is completely dephasing.   The evolution under
this channel,  which may be  called the maximally  dephasing component
(MDC) channel, is given by Eq.  (\ref{eq:diag}).  Because
all Kraus operators of this channel are of rank 1, it has the property
of   being  \textit{entanglement   breaking}.  In  any
dimension,   it  has   an   analytical  operator   sum  (as   against,
sum-difference) representation.

A quantum  channel $\mathcal{E}$  is called entanglement  breaking and
trace-preserving (EBT)  if given any  input state $\Gamma$,  the state
$(\mathbb{I}  \otimes  \mathcal{E})\Gamma$   is  separable.   This  is
equivalent    to    the    separability    of    the    Choi    matrix
$(I\otimes\mathcal{E})|\psi^+\rangle\langle\psi^+|$  and  also to  the
condition that it can be expressed in the \textit{Holevo form}
\begin{equation}
\mathcal{E}_{EB}(\rho)=\sum_iR_iTr(F_i\rho),  
\label{eq:holevo}
\end{equation}
where $\{R_i\}$ is  a set of density operators and  $\{F_i\}$ is a set
of positive operator valued measures (POVM) \cite{hsr03}.

The  2AD  channel is  asymptotically  ($t\longrightarrow \infty$)  EBT
because only the Kraus operators  elements of the MDC survive.  In the
asymptotic  limit the  channel is  characterized by  the fact  that in
(\ref{eq:pma}), $H, F,  G \longrightarrow 1$, whereas $A,  B, C, D, E,
J,  K, L,  M, P,  Q,  X, Y  \longrightarrow 0$.   Therefore, only  the
following  4 rank-1  Kraus operators,  which are  elements of  the MDC
channel,  survive:  \be  K_H^+(\infty)  =  \left(  \begin{array}{cccc}
  0&0&0&0\\ 0&0&0&0\\ 0&0&0&0\\ 1&0&0&0
\end{array}\right);\hspace{0.25cm}
K_F^+(\infty) = \begin{pmatrix}
0&0&0&0\\
0&0&0&0\\
0&0&0&0\\
0&1&0&0
\end{pmatrix};\hspace{0.25cm}
K_G^+(\infty) = \begin{pmatrix}
0&0&0&0\\
0&0&0&0\\
0&0&0&0\\
0&0&1&0
\end{pmatrix}; \hspace{0.25cm}
K_1^+(\infty) = \begin{pmatrix}
0&0&0&0\\
0&0&0&0\\
0&0&0&0\\
0&0&0&1
\end{pmatrix}.
\label{eq:kinf}
\ee  This channel produces  the asymptotic  state $|\Psi_\infty\rangle
\equiv  |11\rangle$, so  that the  corresponding Choi  matrix  has the
separable  form  $|11\rangle\langle11|\otimes I$  and  the channel  is
entanglement breaking: for an arbitrary initial state of the 2-qubit
system possibly entangled with any other system, asymptotically
the 2-qubit system factors out to $|11\rangle$.

Furthermore, asymptotically  it is of an extreme  point of entanglement
breaking, whereby the channel maps  all input states to a single point
in state space,  the pure state $|11\rangle$: it is  thus a {\it point
  channel}  \cite{hsr03}.  The Kraus  operators of  EB channel  can be
expressed  as  $K_i=\sqrt{R_i}|jk\rangle\langle lm|\sqrt{F_i}$,  which
are seen  to reproduce the  expression in Eq.  (\ref{eq:holevo}).  For
the    considered    point    channel    $R_i=R=|11\rangle\langle11|$,
$F_i=\{|00\rangle\langle00|,|01\rangle\langle01|,|10\rangle\langle10|
,|11\rangle\langle11|\}$ and  the resulting  $K_i$ are easily  seen to
conicide with the operators in Eq. (\ref{eq:kinf}).

A special case of the entanglement breaking channel is the
quantum-to-classical (QC) measurement map, wherein
the states $R_i$ in Eq. (\ref{eq:holevo})
are orthogonal projectors $|e_j\rangle\langle e_j|$
of a fixed basis. Acting on any 2-system state, it produces
a QC state, which has the form:
\begin{equation}
\rho^{QC} = \sum_j p_j\sigma_j \otimes
|e_j\rangle\langle e_j|.
\label{eq:qc}
\end{equation}
A fundamental result here, which refines the
equivalent result for EBT channels, is that
a channel $\Lambda$ is of QC-type (meaning that $(\mathbb{I}
\otimes \Lambda)(\rho_{AB})$ is a QC state for any
bipartite state $\rho_{AB}$) if and only if the correponding Choi 
matrix is a QC state \cite{korbicz}.  
A particular feature of interest here is that, 
for any QC-type channel $\Lambda$, given 
any orthonormal basis $\{\phi_j\}$, there exists at least one state
$\rho^\ast(\phi)$ diagonal in this basis, which is 
$N$-copy spectrum-broadcastable using the channel.
If $\phi$ is identified with the channel basis $\{|e_j\rangle\}$,
then $\rho^\ast(e)$ is fully broadcastable. Here the broadcast is
state $\sigma$ shared between Alice and $N$ other parties,
such that the reduced density matrix at each party 
has the same eigenvalue spectrum as $\rho^\ast(\phi)$
(spectrum broadcast) or is identical with $\rho^\ast(\phi)$
(full broadcast).

The question then arises of whether the MDC channel is also
of QC-type. The answer, according the above quoted result, is
in the negative, as seen from the matrix $\mathcal{B}_{\rm diag}$,
formed by the diagonal terms of
$\mathcal{B}$ in Eq. (\ref{eq:pma}), which cannot be cast
in the form (\ref{eq:holevo}) with $R_i$ given by
orthogonal projectors. To see this, we note that because only
diagonal terms are present, both $\sigma_j$ and
$|e_j\rangle$ in Eq. (\ref{eq:qc}) must be diagonal in the
computational basis. Then it is clear why 
$\mathcal{B}_{\rm diag}$, though separable, is not QC. 
In particular, looking at the first 4 diagonal terms, 
we should have $C=H=0$, which is not in general guaranteed.

\subsection{The purely dephasing component}

By  contrast,  a non-dissipative  and  purely  dephasing operation  is
obtained  from  a  channel  comprising the  extended  Kraus  operators
$K_{j}^{\pm}$ $j \in  \textbf{S} \equiv \{J,M,L,P,T,U,V,Q,R,S \}$. This has
the effect  of dephasing the  off-diagonal terms, but killing  off the
diagonal terms. Restoration of  the diagonal components is obtained by
use of positive Kraus operators  given by all the projetors: $\Pi_{00}
\equiv |00\rangle\langle 00|$, $\Pi_{01}\equiv |01\rangle\langle 01|$,
$\Pi_{10}\equiv    |10\rangle\langle    10|$    and    $\Pi_{11}\equiv
|11\rangle\langle    11|$,   which    collectively   form    the   set
$\textbf{U}$. The  elements of this  set are just the  Kraus operators
$K_D^+, K_A^+,  K_B^+$ and $K_1^+$ in  Eq. (\ref{diagonalkraus}), with
$D, A, B = 1$.
 
Together  the  operators  in  the  set  $\textbf{S}  \cup  \textbf{U}$
constitute a CP map that is  a purely dephasing component (PDC) of the
2AD  channel,  which  leaves  the  populations  (diagonal  components)
unchanged, but otherwise reproduces the (dephasing) effects of the 2AD
channel. Its  action is given  by: 
\bea 
\rho^\prime_{\rm pd}  &=& 
 \sum_{j  \in \textbf{U}} K_j^+\rho(K_j^+)^\dag 
  + \sum_{j  \in \textbf{S}} K_j^{\pm}\rho(K_j^\pm)^\dag \nonumber\\
&=& \left(
\begin{array}{clclr}
\rho_{ee}  &J\rho_{es}&M\rho_{ea}&L\rho_{eg}\\
J^\ast\rho_{se}&\rho_{ss} &P\rho_{sa}&T\rho_{sg}+(U+iV)\rho_{es}\\
M^\ast\rho_{ae} & P^\ast\rho_{as}&\rho_{aa}&Q\rho_{ag}+(iS-R)\rho_{ea}\\
L^\ast\rho_{ge} & T^\ast\rho_{gs}+(U^\ast - i V^\ast)\rho_{se}
&Q^\ast\rho_{ga}+ (-iS^\ast - R^\ast)\rho_{ae}&
\rho_{gg}
\end{array}
\right).
\label{eq:pdp}
\eea
Unlike MDC, this is not entanglement breaking at finite time.
To see this, suppose that the input state is
$\psi_{\rm in} \equiv \frac{1}{\sqrt{2}}(|00,00\rangle + |11,11\rangle)$,
i.e., a 1-bit entanglement between the given 2-qubit system,
with another 2-qubit system. The effect of the MDC channel
is seen to be
$$\rho_{\rm out} =
\frac{1}{2}\left(|00,00\rangle\langle 00,00| +
	|11,11\rangle\langle 11,11| +
	J(e^{i\phi_L}|00,00\rangle\langle 11,11| +
	e^{-i\phi_L}|11,11\rangle\langle 00,00|)\right).
$$
The above state lives in four dimensional Hilbert space, given
by $\mathcal{H}_2 \otimes \mathcal{H}_2$, spanned by kets
$\{|00\rangle, |11\rangle\}$ in each two-qubit system.
The state is thus entirely equivalent to a correlated state of
two qubits given by:
$$\rho_{\rm out}^\prime =
\frac{1}{2}\left(|0,0\rangle\langle 0,0| +
	|1,1\rangle\langle 1,1| +
	J(e^{i\phi_L}|0,0\rangle\langle 1,1| +
	e^{-i\phi_L}|1,1\rangle\langle 0,0|)\right),
$$ 
for  which the concurrence \cite{ckw00}, a  measure of entanglement
        for a mixed 2-qubit  system is $J$, implying that entanglement
        is not broken except  asymptotically (when $J \rightarrow 0$).

\section{Conclusions}

In the problem  of deriving the Kraus representation  of a given noise
dynamics  via  the Choi-Jamiolkowski  approach,  we  have developed  a
method  that  circumvents the  impasse  due  to  the Abel-Galois  no-go
theorem  for  the  algebraic  solution  to quintic  and  higher  order
polynomials. Our  idea is to  obtain a Hermitian decomposition  of the
Choi  matrix, which  yield a  set of  `positive' and  `negative' Kraus
operators, satisfying the completeness condition. 

The price to  pay is that, in general, the  sufficient number of Kraus
operators  for a  $d$-dimensional  system is  $d^4$,  rather than  the
usually sufficient  number of $d^2$. 

We have  applied this formalism to three  problems: the sum-difference
representation of the 2-qubit  amplitude damping channel; defining two
mathematically interesting  limits of any channel  in this representation,
namely  the MDC  and  PDC channels,  whose  entanglement breaking  
property  was studied. In particular, the  MDC channel is
EBT, whereas  the PDC channel  is not, except
asymptotically.

\bibliography{openqip}

\appendix

\section{The two-qubit amplitude damping channel\label{sec:a}}

Information useful for the description of the 2AD  channel, discussed in Sec. IV, are:

$A=e^{-2\Gamma t}$; 
$B=e^{-(\Gamma +\Gamma_{12})t}$;
$C=\frac{\Gamma +\Gamma_{12}}{\Gamma -\Gamma_{12}}(1-e^{-(\Gamma -\Gamma_{12})t}) e^{-(\Gamma +\Gamma_{12})t}$;
$D=e^{-(\Gamma -\Gamma_{12})t}$;
$E=\frac{\Gamma -\Gamma_{12}}{\Gamma +\Gamma_{12}}(1-e^{-(\Gamma +\Gamma_{12})t})e^{-(\Gamma -\Gamma_{12})t}$;
$F=1-e^{-(\Gamma +\Gamma_{12})t}$;
$G=1-e^{-(\Gamma -\Gamma_{12})t}$;
$J=e^{-i(\omega_0 -\Omega_{12})t} e^{(3\Gamma  +\Gamma_{12})t/2}$;
$L=e^{-i2\omega_0 t}e^{-\Gamma  t}$;
$M=e^{-i(\omega_0 +\Omega_{12})t} e^{-(3\Gamma  -\Gamma_{12})t/2}$;
$P=e^{-i2\Omega_{12}t}e^{-\Gamma  t}$;
$Q=e^{-i(\omega_0 -\Omega_{12})t} e^{-(\Gamma  -\Gamma_{12})t/2}$;
$T=e^{-i(\omega_0 +\Omega_{12})t} e^{-(\Gamma +\Gamma_{12})t/2}$;
\begin{eqnarray}
H&=&\frac{\Gamma +\Gamma_{12}}{2\Gamma }\left[1-\frac{2}{\Gamma -\Gamma_{12}}\left(\frac{\Gamma +\Gamma_{12}}{2}(1-e^{-(\Gamma -\Gamma_{12}t})
+\frac{\Gamma -\Gamma_{12}}{2}\right)e^{-(\Gamma +\Gamma_{12})t}\right]\\ \nonumber
&+&\frac{\Gamma -\Gamma_{12}}{\Gamma +\Gamma_{12}}\left[(1-e^{-(\Gamma -\Gamma_{12})t})-\frac{\Gamma -\Gamma_{12}}{2\Gamma }(1-e^{-2\Gamma t})\right],\\ \nonumber
R&=&\frac{\Gamma -\Gamma_{12}}{\Gamma ^2+4\Omega_{12}^2} e^{-i(\omega_0 -\Omega_{12})t}  e^{-(\Gamma -\Gamma_{12})t/2}
\left[2\Omega_{12} e^{-\Gamma t} \sin(2\Omega_{12} t)+\Gamma (1-e^{-\Gamma t}\cos(2\Omega_{12} t))\right], \nonumber \\
S&=&\frac{\Gamma -\Gamma_{12}}{\Gamma ^2+4\Omega_{12}^2} e^{-i(\omega_0 -\Omega_{12})t}  e^{-(\Gamma -\Gamma_{12})t/2}
\left[2\Omega_{12}(1-e^{-\Gamma t}\cos(2\Omega_{12} t)) -\Gamma e^{-\Gamma  t} \sin(2\Omega_{12} t) )\right], \nonumber \\
U&=&\frac{\Gamma +\Gamma_{12}}{\Gamma ^2+4\Omega_{12}^2} e^{-i(\omega_0 + \Omega_{12})t}  e^{-(\Gamma +\Gamma_{12})t/2}
\left[2\Omega_{12} e^{-\Gamma  t} \sin(2\Omega_{12} t) + \Gamma (1-e^{-\Gamma t}\cos(2\Omega_{12} t))\right],\\ \nonumber
V&=&\frac{\Gamma +\Gamma_{12}}{\Gamma ^2+4\Omega_{12}^2} e^{-i(\omega_0 +\Omega_{12})t}  e^{-(\Gamma +\Gamma_{12})t/2}
\left[2\Omega_{12}(1-e^{-\Gamma t}\cos(2\Omega_{12} t)) -\Gamma e^{-\Gamma  t} \sin(2\Omega_{12} t) )\right].
\end{eqnarray}
The terms $\omega_0$, $\Omega_{12}$, $\Gamma$, $\Gamma_{12}$ are as defined in \cite{brs2}.

\end{document}